\documentclass{igaia}

\usepackage[dvips]{graphicx}

\theoremstyle{definition}

\theoremstyle{remark}

\newcommand{\ri}{{\rm i}}
\newcommand{\rd}{{\rm d}}

\begin{document}
\title[Quantum states and space-time causality]{
Quantum states and space-time causality}

\author[D.~C.~Brody]{Dorje C. Brody}
\address{Blackett Laboratory, Imperial College, London
SW7 2BZ, UK} \email{dorje@imperial.ac.uk}
\urladdr{http://www.imperial.ac.uk/people/d.brody}

\author[L.~P.~Hughston]{Lane P. Hughston}
\address{Department of Mathematics, King's
College London, London WC2R 2LS, UK} \email{lane.hughston@kcl.ac.uk}
\urladdr{http://www.mth.kcl.ac.uk/staff/l$\_$hughston.html}

\maketitle

\begin{abstract}
Space-time symmetries and internal quantum symmetries can be
placed on equal footing in a hyperspin geometry. Four-dimensional
classical space-time emerges as a result of a decoherence that
disentangles the quantum and the space-time degrees of freedom. A
map from the quantum space-time to classical space-time that preserves the
causality relations of space-time events is necessarily a density
matrix.
\end{abstract}

\section{Introduction}

This article presents a programme for the unification of space-time
and internal quantum symmetries.  An important role is played in this theory
by certain higher-dimensional analogues of spinors. In
four-dimensional space-time there is a local isomorphism between the
Lorentz group $SO(1,3)$ and the spin transformation group
$SL(2,{\mathbb C})$. In higher dimensions, however, this relation
breaks down and we are left with two concepts of spinors---one for
the groups $SO(N,{\mathbb C})$, and one for the groups
$SL(r,{\mathbb C})$. The spinors associated with $SO(N,{\mathbb C})$
are the so-called Cartan spinors. The study of Cartan spinors has a
long history, and there is a beautiful geometry
associated with these spinors. The spinors associated with
$SL(r,{\mathbb C})$, called `hyperspinors', have the advantage of
being more directly linked with quantum mechanics. In fact, a
relativistic model for hyperspin arises when one considers
`multiplets' of two-component spinors, i.e. expressions of the form
$\xi^{{\bf A}i}$ and $\eta^{{\bf A}^\prime}_i$, where ${\bf A},{\bf
A}'$ are standard spinor indices and $i=1,2, \ldots,n$ is an
`internal' index. In the general case ($n=\infty$) we can think of
$\xi^{{\bf A}i}$ as an element of the tensor product space ${\mathbb
S}^{{\bf A}i} ={\mathbb S}^{\bf A}\otimes{\mathbb H}^i$, where
${\mathbb S}^{\bf A}$ is the space of two-component
spinors, and ${\mathbb H}^i$ is an infinite-dimensional complex
Hilbert space.

The theory of hyperspin constitutes a natural starting place for
building up a theory of quantum geometry or, as we shall call it
here, \emph{quantum space-time}. The hyperspinor route has the
virtue that the resulting higher-dimensional space-time has a rich
causal structure associated with it, and as a consequence is
well-positioned to form the geometrical basis of a physical theory.

\section{Relativistic causality}

Let us review briefly the role of two-component spinors in the
description of four-dimensional Minkowski space. We use bold
upright Roman letters to denote two-component spinor indices, and we
adopt the standard conventions for the algebra of two-component
spinors~\cite{penrose}. Then we have the following correspondence
between two-by-two Hermitian matrices $x^{{\bf AA}^\prime}~({\bf
A},{\bf A}^\prime=1,2)$ and the positions $x^{\rm a}~({\rm
a}=0,1,2,3)$ of space-time points relative to some origin. More
explicitly, in a standard basis this correspondence is given by
\begin{eqnarray}
\frac{1}{\sqrt{2}} \left( \begin{array}{ll} t+z & x+\ri y \\ x-\ri
y & t-z \end{array} \right) \quad \longleftrightarrow \quad
(t,x,y,z). \label{eq:2}
\end{eqnarray}
We thus obtain the fundamental relation $2\det(x^{{\bf AA}^\prime})
= t^2-x^2-y^2-z^2$. It follows that two-component spinors are
connected both with quantum mechanics and with the causal structure
of space-time. It is a peculiar aspect of relativistic physics that
there is a link of this nature between the spin degrees of freedom
of spin one-half particles, and the causal geometry of
four-dimensional space-time.

For the interval between a pair of points $x^{{\bf AA}^\prime}$ and
$y^{{\bf AA}^\prime}$ in space-time we write $r^{{\bf AA}^\prime} =
x^{{\bf AA}^\prime}-y^{{\bf AA}^\prime}$. It follows that
$2\det(r^{{\bf AA}^\prime}) = \epsilon_{\bf AB}\epsilon_{{\bf
A}^\prime{\bf B}^\prime} r^{{\bf AA}^\prime} r^{{\bf BB}^\prime}$,
where $\epsilon_{\bf AB}$ is the antisymmetric spinor. If we adopt
the `index clumping' convention and write ${\rm a}={\bf AA}'$, ${\rm
b}={\bf BB}'$, and so on, whereby a pair of spinor indices, one
primed and the other unprimed, corresponds to a space-time vector
index, then we can write $\epsilon_{\bf AB}\epsilon_{{\bf
A}^\prime{\bf B}^\prime} r^{{\bf AA}^\prime} r^{{\bf BB}^\prime} =
g_{\rm ab} r^{\rm a} r^{\rm b}$ for the corresponding squared
space-time interval. Thus we identify $g_{\rm ab} = \epsilon_{\bf
AB}\epsilon_{{\bf A}^\prime{\bf B}^\prime}$ as the metric of
Minkowski space.

There are three different situations that can arise for the interval
$r^{\rm a}$. The first case is $g_{\rm ab}r^{\rm b}=0$; the second
case is $g_{\rm ab}r^{\rm b}\neq0$ and $g_{\rm ab}r^{\rm a}r^{\rm
b}=0$; and the third case is $g_{\rm ab}r^{\rm a}r^{\rm b}\neq0$.
Each of these cases gives rise to a canonical form for the interval
$r^{{\bf AA}^\prime}$, with various sub-cases, which can be
summarised as follows: (i) $g_{\rm ab}r^{\rm b}=0$ implies $r^{{\bf
AA}^\prime}=0$ (zero separation); (ii) $g_{ab} r^a r^b=0$ implies
$r^{{\bf AA}^\prime}=\alpha^{\bf A} {\bar\alpha}^{{\bf A}^\prime}$
(future-pointing null separation) or $r^{{\bf
AA}^\prime}=-\alpha^{\bf A} {\bar\alpha}^{{\bf A}^\prime}$
(past-pointing null separation); (iii) $g_{ab} r^a r^b\neq0$ implies
$r^{{\bf AA}^\prime}= \alpha^{\bf A} {\bar\alpha}^{{\bf A}^\prime} +
\beta^{\bf A}{\bar\beta}^{{\bf A}^\prime}$ (future-pointing
time-like separation), $r^{{\bf AA}^\prime} = \alpha^{\bf A}
{\bar\alpha}^{{\bf A}^\prime} - \beta^{\bf A}{\bar\beta}^{{\bf
A}^\prime}$ (space-like separation), or $r^{{\bf
AA}^\prime}=-\alpha^{\bf A}{\bar\alpha}^{{\bf A}^\prime} -
\beta^{\bf A}{\bar\beta}^{{\bf A}^\prime}$ (past-pointing time-like
separation). Once the canonical form for $r^{{\bf AA}^\prime}$ is
specified, so is the causal relationship that it determines.

\section{Hyperspin spaces and quantum space-time}

In the case of hyperspinors (introduced by
Finkelstein~\cite{finkelstein1}, cf. also
\cite{finkelstein2,holm,honeycutt}) we replace the two-component
spinors of four-dimensional space-time with $r$-component spinors.
We can regard hyperspin space as the vector space ${\mathbb C}^r$
with some extra structure. In particular, in addition to the primary
hyperspin space we have three other vector spaces---the dual
hyperspin space, the complex-conjugate hyperspin space, and the dual
complex-conjugate hyperspin space.

Let us write ${\mathbb S}^{A}$ and ${\mathbb S}^{A'}$ for the
complex $r$-dimensional vector spaces of unprimed and primed
hyperspinors. For hyperspinors we use italic indices to distinguish
them from the boldface indices used for two-component spinors. We
assume that ${\mathbb S}^{A}$ and ${\mathbb S}^{A'}$ are related by
an anti-linear isomorphism under complex conjugation. Thus if
$\alpha^{A}\in{\mathbb S}^{A}$, then $\alpha^{A}\to
{\bar\alpha}^{A'}$ under complex conjugation, where
${\bar\alpha}^{A'} \in {\mathbb S}^{A'}$. The dual spaces associated
with ${\mathbb S}^{A}$ and ${\mathbb S}^{A'}$ are denoted ${\mathbb
S}_{A}$ and ${\mathbb S}_{A'}$. If $\alpha^{A}\in{\mathbb S}^{A}$
and $\beta_{A}\in{\mathbb S}_{A}$, then their inner product is
denoted $\alpha^{A}\beta_{A}$; if $\gamma^{\,A'}\in {\mathbb
S}^{A'}$ and $\delta_{A'}\in {\mathbb S}_{A'}$ then their inner
product is $\gamma^{\,A'}\delta_{A'}$. We also introduce the totally
antisymmetric hyperspinors of rank $r$ associated with ${\mathbb
S}^{A}$, ${\mathbb S}_{A}$, ${\mathbb S}^{A'}$, and ${\mathbb
S}_{A'}$. These are denoted $\varepsilon^{AB\cdots C}$,
$\varepsilon_{AB\cdots C}$, $\varepsilon^{A'B'\cdots C'}$, and
$\varepsilon_{A'B'\cdots C'}$, and satisfy the relations
$\varepsilon^{AB\cdots C} \varepsilon_{AB\cdots C}=r!$,
$\varepsilon^{A'B'\cdots C'} \varepsilon_{A'B'\cdots C'}=r!$, and
$\varepsilon_{A'B'\cdots C'}= {\bar\varepsilon}_{A'B'\cdots C'}$.

Next we introduce the complex matrix space ${\mathbb C}^{AA'}
={\mathbb S}^{A}\otimes{\mathbb S}^{A'}$. An element
$x^{\,AA'}\in{\mathbb C}^{AA'}$ is said to be \emph{real} if it
satisfies the (weak) Hermitian property $x^{\,AA'}= {\bar x}^{A'A}$.
We shall have more to say about weak versus strong Hermiticity in
connection with the idea of symmetry breaking. We denote the vector
space of real elements of ${\mathbb C}^{AA'}$ by ${\mathbb
R}^{AA'}$. The points of ${\mathbb R}^{AA'}$ constitute what we call
the quantum space-time ${\mathcal H}^{r^2}$ of dimension $r^2$. We
regard $\mathcal{C\!H}^{r^2}={\mathbb C}^{AA'}$ as the
complexification of ${\mathcal H}^{r^2}$. Many problems in
${\mathcal H}^{r^2}$ are best first approached as problems in
$\mathcal{C\!H}^{r^2}$.

Let $x^{\,AA'}$ and $y^{\,AA'}$ be points in ${\mathcal H}^{r^2}$,
and write $r^{\,AA'}=x^{\,AA'} -y^{\,AA'}$ for their separation
vector, which is independent of the choice of origin. Using the
index-clumping convention we set $x^{\rm a}=x^{\,AA'}$, $y^{\rm
a}=y^{\,AA'}$, $r^{\rm a} = r^{\,AA'}$, and for the separation of
$x^{\rm a}$ and $y^{\rm a}$ in ${\mathcal H}^{r^2}$ we write $r^{\rm
a}=x^{\rm a}- y^{\rm a}$. There is a natural causal structure
induced on such intervals by Finkelstein's \emph{chronometric
tensor}~\cite{finkelstein1}, defined by the relation $g_{{\rm
ab\cdots c}} = \varepsilon_{AB\cdots C}\, \varepsilon_{A'B'\cdots
C'}$. The chronometric tensor is of rank $r$, is totally symmetric
and is nondegenerate in the sense that $v^{\rm a}g_{\rm ab \cdots
c}\neq0$ for any vector $v^{\rm a}\neq0$. We say that $x^{\rm a}$
and $y^{\rm a}$ in ${\mathcal H}^{r^2}$ have a `degenerate'
separation if the chronometric form $\Delta(r)=g_{{\rm ab\cdots
c}}r^{\rm a}r^{\rm b}\cdots r^{\rm c}$ vanishes for $r^{\rm
a}=x^{\rm a}- y^{\rm a}$. Degenerate separation is equivalent to the
vanishing of the determinant of $r^{\,AA'}$.

If the hyperspin space has dimension $r=2$, this condition reduces
to the case where $x^{\rm a}$ and $y^{\rm a}$ are null-separated in
Minkowski space. For $r>2$ the situation is more complicated since
various degrees of degeneracy can arise between two points of a
quantum space-time. In the case $r=3$, for example, the quantum
space-time has dimension nine, and the chronometric form is
$\Delta=g_{\rm abc}r^{\rm a}r^{\rm b}r^{\rm c}$. The different
possibilities that can arise for the separation vector are as
follows: (i) $g_{\rm abc}r^{\rm c}=0$ implies $r^{\,AA'}=0$ (zero
separation); (ii) $g_{\rm abc}r^{\rm b}r^{\rm c}=0$ and $g_{\rm abc}
r^{\rm c}\neq0$ implies $r^{\,AA'}= \alpha^A {\bar\alpha}^{A'}$
(future-pointing null separation) or $r^{\,AA'} =-\alpha^A
{\bar\alpha}^{A'}$ (past-pointing null separation); (iii) $\Delta=0$
and $g_{\rm abc}r^{\rm b}r^{\rm c} \neq 0$ implies
$r^{\,AA'}=\alpha^A{\bar\alpha}^{A'} + \beta^A{\bar\beta}^{A'}$
(degenerate time-like future-pointing separation), $r^{\,AA'} =
\alpha^A{\bar\alpha}^{A'} - \beta^A {\bar\beta}^{A'}$ (degenerate
space-like separation), or $r^{\,AA'}=-\alpha^A{\bar\alpha}^{A'} -
\beta^A{\bar\beta}^{A'}$ (degenerate time-like past-pointing
separation); (iv) $\Delta\neq0$ and $g_{\rm ab} r^{\rm a}r^{\rm
b}\neq0$ implies $r^{\,AA'}=\alpha^A{\bar\alpha}^{A'} +
\beta^A{\bar\beta}^{A'} + \gamma^{\,A}{\bar\gamma}^{\,A'}$
(future-pointing time-like separation), $r^{\,AA'} = \alpha^A
{\bar\alpha}^{A'} + \beta^A {\bar\beta}^{A'} - \gamma^{\,A}
{\bar\gamma}^{\,A'}$ (future semi-space-like separation),
$r^{\,AA'}=\alpha^A {\bar\alpha}^{A'} - \beta^A{\bar\beta}^{A'}-
\gamma^{\,A} {\bar\gamma}^{\,A'}$ (past semi-space-like separation),
or $r^{\,AA'}=-\alpha^A {\bar\alpha}^{A'} - \beta^A{\bar\beta}^{A'}-
\gamma^{\,A} {\bar\gamma}^{\,A'}$ (past-pointing time-like
separation).

When the separation of two points of a quantum space-time is
degenerate, we define the `degree' of degeneracy by the rank of the
matrix $r^{\,AA'}$. Null separation is the case for which the
degeneracy is of the first degree, i.e. $r^{\,AA'}$ is of rank one,
and thus satisfies a system of quadratic relations of the form
$g_{{\rm ab\cdots c}}r^{\rm a}r^{\rm b}=0$. This implies
$r^{\,AA'}=\pm \alpha^{A}{\bar\alpha}^{A'}$ for some $\alpha^A$. In
the case of degeneracy of the second degree, $r^{\,AA'}$ is of rank
two and satisfies a set of cubic relations given by $g_{{\rm
abc\cdots d}} r^{\rm a} r^{\rm b}r^{\rm c}=0$. In this situation
$r^{\,AA'}$ can be put into one of the following canonical forms:
(a) $r^{\,AA'}=\alpha^{A} {\bar\alpha}^{A'}+
\beta^A{\bar\beta}^{A'}$, (b) $r^{\,AA'}= \alpha^A{\bar\alpha}^{A'}-
\beta^{A} {\bar\beta}^{A'}$, and (c) $r^{\,AA'}=
-\alpha^{A}{\bar\alpha}^{A'}- \beta^{A} {\bar\beta}^{A'}$. In case
(a), the point $x^{\rm a}$ lies to the future of the point $y^{\rm
a}$, and $r^{\rm a}$ can be thought of as a degenerate
future-pointing time-like vector. In case (b), $r^{\rm a}$ is a
degenerate space-like vector. In case (c), $x^{\rm a}$ lies to the
past of $y^{\rm a}$, and $r^{\rm a}$ is a degenerate past-pointing
time-like vector. A similar analysis can be applied to degenerate
separations of other intermediate degrees.

If the determinant of the $r$-by-$r$ weakly Hermitian matrix
$r^{\,AA'}$ is nonvanishing, and $r^{\,AA'}$ is thus of maximal
rank, then the chronometric form is nonvanishing. In that case
$r^{\,AA'}$ can be represented in the following canonical form:
\begin{eqnarray}
r^{\,AA'}=\pm \alpha^{A}{\bar\alpha}^{A'} \pm \beta^{A}
{\bar\beta}^{A'} \pm \cdots \pm \gamma^{\,A}{\bar\gamma}^{\,A'},
\end{eqnarray}
with the presence of $r$ nonvanishing terms, where the $r$
hyperspinors $\alpha^{A},\beta^{A}, \cdots, \gamma^{\,A}$ are
linearly independent. Let us write $(p,q)$ for the numbers of plus
and minus signs appearing in the canonical form for $r^{\,AA'}$. We
call $(p,q)$ the `signature' of $r^{\,AA'}$. When the signature is
$(r,0)$ or $(0,r)$ we say that $r^{\,AA'}$ is future-pointing
time-like or past-pointing time-like, respectively. Then we define
the proper time interval between the events $x^{\rm a}$ and $y^{\rm
a}$ by the formula $\|x-y\| = |\Delta|^{1/r}$. In the case $r=2$ we
recover the Minkowskian proper-time interval.

A remarkable feature of the causal structure of a quantum space-time
is that many of the physical features of the causal structure of
Minkowski space are preserved. In particular, the space of
future-pointing time-like vectors forms a convex cone. The same is
true for the structure of the associated momentum space, from which
it follows that we can give a good definition of `positive energy'.

\section{Equations of motion}

Let $\lambda\mapsto x^{\,AA'}(\lambda)$ define a smooth curve
$\gamma$ in ${\mathcal H}^{r^2}$ for $\lambda \in [a,b]
\subset{\mathbb R}$. Then $\gamma$ is said to be time-like if the
tangent vector $v^{AA'}(\lambda) = \rd x^{\,AA'}(\lambda)
/\rd\lambda$ is time-like and future-pointing. In that case we
define the proper time $s$ elapsed along $\gamma$ by
\begin{eqnarray}
s = \int_a^b \left[ g_{{\rm ab}\cdots{\rm c}} v^{{\rm a}} v^{\rm
b}\cdots v^{\rm c} \right]^{1/r} \rd \lambda . \label{eq:21}
\end{eqnarray}
In the case of a very small time interval, we can write this in the
pseudo-Finslerian form $(\rd s)^r = g_{{\rm ab}\cdots{\rm c}}\rd
x^{\rm a} \rd x^{\rm b} \cdots \rd x^{\rm c}$. For $r=2$ this
reduces to the standard pseudo-Riemannian expression for the line
element.

Now consider the condition $\gamma$ must satisfy in order to be a
geodesic in ${\mathcal H}^{r^2}$. In the case of a time-like curve,
we can choose the proper time as the parameter along the curve. The
equation of motion for a time-like geodesic is
obtained by an application of the calculus of variations to formula
(\ref{eq:21}). We assume the variation vanishes at the endpoints.
Writing $L$ for the integrand in (\ref{eq:21}), we can use a
standard argument to show that $x^{\rm a} (s)$ describes a geodesic
only if the velocity vector $v^{\rm a}$ satisfies the Euler-Lagrange
equation $\rd(\partial L/\partial v^{\rm a})/\rd s= 0$. It follows
that if $y^{\rm a}$ and $z^{\rm a}$ are quantum space-time points
such that $y^{\rm a}-z^{\rm a}$ is time-like and future-pointing,
then the affinely parametrised geodesic $\gamma$ connecting these
points in ${\mathcal H}^{r^2}$ is given by (cf.
Busemann~\cite{busemann})
\begin{eqnarray}
x^{\rm a}(s) = z^{\rm a} + \frac{y^{\rm a}-z^{\rm a}}{[\Delta
(y,z)]^{1/r}}\, s,
\end{eqnarray}
for $s\in(-\infty,\infty)$, where $\Delta(y,z)=g_{\rm ab\cdots
c}(y^{\rm a}-z^{\rm a})(y^{\rm b}-z^{\rm b})\cdots (y^{\rm
c}-z^{\rm c})$.

\section{Conserved quantities}

The chronometric form for the separation between two points is
invariant when the points of ${\mathcal H}^{r^2}$ are subjected to
transformations of the form
\begin{eqnarray}
x^{\,AA'} \to \lambda^{A}_{B} {\bar\lambda}^{A'}_{B'} x^{BB'} +
\beta^{AA'} . \label{eq:33}
\end{eqnarray}
Here $\beta^{AA'}$ is a translation in the quantum space-time, and
$\lambda^{A}_{B}$ is an element of $SL(r,{\mathbb C})$. The relation
of this group of transformations to the Poincar\'e group in the case
$r=2$ is clear. Indeed, one of the attractive features of the
extension of space-time geometry that we are presenting is that the
hyper-Poincar\'e group admits such a description, which has a number
of important physical consequences.

More generally, the proper hyper-Poincar\'e group preserves the
signature of any space-time interval, whether or not the interval is
degenerate, and hence preserves the causal relations between events.
If $L^{\rm a}_{\rm b}= \lambda^{A}_{B} {\bar\lambda}^{A'}_{B'}$ for
some $\lambda^{A}_{B} \in SL(r,{\mathbb C})$, we refer to a map of
the form $r^{\rm a}\to L^{\rm a}_{\rm b}r^{\rm b}$ as a
\emph{hyper-Lorentz transformation}.

The real dimension of the hyper-Lorentz group is $2r^2-2$, and hence
the real dimension of the hyper-Poincar\'e group is $3r^2-2$. The
dimension of the hyper-Poincar\'e group thus grows linearly with the
dimension of the quantum space-time. This can be contrasted with the
dimension of the group arising if we endow ${\mathbb R}^{r^2}$ with
a Lorentzian metric with signature $(1,r^2-1)$. In that case the
associated pseudo-orthogonal group has dimension $\frac{1}{2}
r^2(r^2-1)$, which together with the translation group gives a total
dimension of $\frac{1}{2}r^2(r^2+1)$. The comparatively low
dimensionality of the hyper-Poincar\'e group arises from the fact
that it preserves a more elaborate system of causal relations than
what one has in the Lorentzian case.

In Minkowski space the symmetries of the Poincar\'e group are
associated with a ten-parameter family of Killing vectors. Thus for
$r=2$ we have the Minkowski metric $g_{\rm ab}$, and the Poincar\'e
group is generated by vector fields $\xi^{\rm a}$ on ${\mathcal
H}^{4}$ that satisfy ${\mathcal L}_\xi g_{\rm ab}=0$, where
${\mathcal L}_\xi$ denotes the Lie derivative. For any vector field
$\xi^{\rm a}$ and any symmetric tensor field $g_{\rm ab}$ we have
${\mathcal L}_\xi g_{\rm ab} = \xi^{\rm c} \nabla_{\!\rm c}g_{\rm
ab} + 2g_{{\rm c}({\rm a}} \nabla_{\!{\rm b})} \xi^{\rm c}$. If
$g_{\rm ab}$ is the metric and $\nabla_{\!\rm a}$ denotes the
torsion-free covariant derivative satisfying $\nabla_{\!\rm a}g_{\rm
bc}=0$, we obtain the Killing equation $\nabla_{\!({\rm a}}\xi_{{\rm
b})}=0$, where $\xi_{\rm a}=g_{\rm ab}\xi^{\rm b}$. The condition
${\mathcal L}_\xi g_{\rm ab}=0$ therefore implies that $\xi^{\rm a}$
is a Killing vector.

For $r>2$ the usual relations between symmetries and Killing vectors
are lost. Instead, we obtain a system of higher-rank Killing
tensors. More specifically, to generate a symmetry of the quantum
space-time the vector field $\xi^{\rm a}$ has to satisfy ${\mathcal
L}_\xi g_{\rm ab\cdots c} = 0$, where $g_{\rm ab\cdots c}$ is the
chronometric tensor. For a vector field $\xi^{\rm a}$ and a
symmetric tensor field $g_{\rm ab\cdots c}$ we have
\begin{eqnarray}
{\mathcal L}_\xi g_{\rm ab\cdots c} = \xi^{\rm d} \nabla_{\!\rm d}
g_{\rm ab\cdots c} + r\,g_{{\rm d}({\rm a\cdots b}} \nabla_{\!{\rm
c})} \xi^{\rm d} .
\end{eqnarray}
In the case of the quantum space-time ${\mathcal H}^{r^2}$ we let
$\nabla_{\!{\rm a}}$ be the natural flat connection for which
$\nabla_{\!{\rm a}} g_{\rm bc\cdots d}=0$. Then to generate a
symmetry of the chronometric structure of ${\mathcal H}^{r^2}$ the
vector field $\xi^{\rm a}$ has to satisfy $g_{{\rm d}({\rm a\cdots
b}} \nabla_{\!{\rm c})} \xi^{\rm d} = 0$. This equation can be
written in a suggestive form if we define a symmetric tensor $K_{\rm
ab\cdots c}$ of rank $r-1$ by setting $K_{\rm ab\cdots c} = g_{\rm
ab\cdots cd}\xi^{\rm d}$. Then it follows that $K_{\rm ab\cdots c}$
satisfies the conditions for a symmetric Killing tensor:
$\nabla_{\!({\rm a}}K_{\rm bc\cdots d)} = 0$. We thus see that
${\mathcal H}^{r^2}$ provides a symmetry group generated by a family
of Killing tensors. The symmetries of the quantum space-time are
generated by a system of $3r^2-2$ irreducible symmetric Killing
tensors of rank $r-1$. The significance of Killing tensors is that
they are associated with conserved quantities. In particular, if the
vector field $v^{\rm a}$ satisfies the geodesic equation, which on a
quantum space-time of dimension $r^2$ is given by $g_{\rm abc\cdots
d}\left( v^{\rm e}\nabla_{\!\rm e}v^{\rm b}\right) v^{\rm c}\cdots
v^{\rm d} = 0$, and if $K_{\rm ab\cdots c}$ is the Killing tensor of
rank $r-1$, then we have the following conservation law: $v^{\rm
e}\nabla_{\!\rm e}\left( K_{\rm ab\cdots c} v^{\rm a} v^{\rm
b}\cdots v^{\rm c}\right) = 0$. In other words, $K_{\rm ab\cdots c}
v^{\rm a}v^{\rm b}\cdots v^{\rm c}$ is a constant of the motion.

\section{Hyper-relativistic mechanics}

In higher-dimensional quantum space-times the conservation laws and
symmetry principles of relativistic physics remain intact. In
particular, the conservation of hyper-relativistic momentum and
angular momentum for a system of interacting particles can be
formulated by use of principles similar to those of the Minkowskian
case. For this purpose we introduce the idea of an `elementary
system' in hyper-relativistic mechanics. Such a system is defined by
its momentum and angular momentum. The hyper-relativistic momentum
of an elementary system is given by a momentum covector $P_{\rm a}$.
The associated mass $m$ is given (cf. \cite{finkelstein1}) by: $m=(
g^{\rm ab\cdots c}P_{\rm a}P_{\rm b}\cdots P_{\rm c})^{1/r}$. The
hyper-relativistic angular momentum is given by a tensor $L^{\rm
b}_{\rm a}$ of the form $L^{\rm b}_{\rm a}=l^{B}_{A}
\delta^{B'}_{A'}+{\bar l}^{B'}_{A'} \delta^{B}_{A}$, where the
hyperspinor $l^{B}_{A}$ is trace-free: $l^{A}_{A}=0$. The angular
momentum is defined with respect to a choice of origin. Under a
change of origin defined by a shift vector $\beta^{\rm a}$ we have
$l^{B}_{A}\to l^B_A + P_{AC'}\beta^{BC'}$. In the case $r=2$ these
formulae reduce to the usual expressions for momentum and angular
momentum in a Minkowskian setting. The real covector $S_{AA'} = \ri
m^{-1} ( l^B_A P_{A'B} - {\bar l}^{B'}_{A'} P_{AB'})$ is invariant
under a change of origin, and can be interpreted as the intrinsic
spin of the elementary system. The magnitude of the spin is
$S=|g^{\rm ab\cdots c} S_{\rm a}S_{\rm b}\cdots S_{\rm c}|^{1/r}$.
In the case of a set of interacting hyper-relativistic systems we
require that the total momentum and angular momentum should be
conserved. This implies conservation of the total mass and spin. We
thus see that the idea of relativistic mechanics carries through to
the case of a general quantum space-time. We shall see later, once
we introduce the idea of symmetry breaking, that hyper-momentum can
be interpreted as the momentum operator for a relativistic quantum
system. Conservation of hyper-momentum then can be thought of as
conservation of four-momentum, in relativistic quantum mechanics, in
the Heisenberg representation.

\section{Weak and strong Hermiticity}

As a prelude to our discussion of symmetry breaking in a quantum
space-time, we digress briefly to review the notions of weak and
strong Hermiticity. This material is relevant to the origin of
unitarity in quantum mechanics. Intuitively speaking, when the weak
Hermiticity condition is imposed on a hyperspinor $x^{\,AA'}$, then
$x^{\,AA'}$ belongs to the real subspace ${\mathbb R}^{AA'}$. The
hyper-relativistic symmetry of a quantum space-time is not affected
by the imposition of this condition. If, however, we break the
symmetry by selecting a preferred time-like direction, then we can
speak of a stronger reality condition whereby \emph{an isomorphism
is established between the primed and unprimed hyperspin spaces}.

We begin with the weak Hermitian property. Let ${\mathbb S}^{A}$
denote, as before, an $r$-dimensional complex vector space. We also
introduce the spaces ${\mathbb S}_{A}$, ${\mathbb S}^{A'}$, and
${\mathbb S}_{A'}$. In general, there is no natural isomorphism
between ${\mathbb S}^{A'}$ and ${\mathbb S}_{A}$, and there is no
natural matrix multiplication law or trace operation defined for
elements of ${\mathbb S}^{A}\otimes{\mathbb S}^{A'}$. Certain matrix
operations are well defined. For example, the determinant of a
generic element $\mu^{AA'}$ is given by $r!\,\det(\mu) =
\varepsilon_{AB\cdots C}\, \varepsilon_{A'B'\cdots C'}\,
\mu^{AA'}\mu^{BB'}\cdots \mu^{CC'}$. The weak Hermitian property is
also well-defined: if ${\bar\mu}^{A'A}$ is the complex conjugate of
$\mu^{AA'}$, then we say that $\mu^{AA'}$ is weakly Hermitian if
$\mu^{AA'}={\bar\mu}^{A'A}$.

Next we consider the strong Hermitian property. In some situations
there may exist a natural map ${\mathbb S}^{A'}\to{\mathbb S}_{A}$
defined by the context of the problem. Such a map is called a
Hermitian correlation. In this case, the complex conjugate of an
element $\alpha^{A}\in{\mathbb S}^{A}$ determines an element
${\bar\alpha}_{A}\in{\mathbb S}_{A}$. For any element
$\mu^{A}_{B}\in {\mathbb S}^{A}\otimes{\mathbb S}_{B}$ we define the
operations of determinant, matrix multiplication, and trace in the
usual manner. The determinant is $r!\, \det(\mu) =
\varepsilon_{AB\cdots C}\, \varepsilon^{PQ\cdots R}\,
\mu^{A}_{P}\mu^{B}_{Q} \cdots \mu^{C}_{R}$, and the Hermitian
conjugate of $\mu^{A}_{\ B}$ is ${\bar\mu}_{A}^{\ B}$. The Hermitian
correlation is given by the choice of a preferred element
$t_{AA'}\in {\mathbb S}_{A} \otimes{\mathbb S}_{A'}$. Then we write
${\bar\alpha}_{A} = t_{AA'}{\bar\alpha}^{A'}$, where
${\bar\alpha}_{A}$ is now called the complex conjugate of
$\alpha^{A}$. When there is a Hermitian correlation ${\mathbb
S}^{A'} \leftrightarrow{\mathbb S}_{A}$, we call the condition
$\mu^{A}_{\ B}= {\bar\mu}^{A}_{\ B}$ the strong Hermitian property.

\section{Symmetry breaking and quantum entanglement}

We proceed to introduce a mechanism for symmetry breaking in a
quantum space-time. We shall make the point that the breaking of
symmetry in a quantum space-time is intimately linked to the notion
of quantum entanglement. According to this point of view, the
introduction of symmetry-breaking in the early stages of the
universe can be understood as a \emph{sequence of phase
transitions}, the ultimate consequence of which is an approximate
disentanglement of a four-dimensional `classical' space-time.

The breaking of symmetry is represented by an `index decomposition'.
In particular, if the dimension $r$ of the hyperspin space is not a
prime number, then a natural method of breaking the symmetry arises
by consideration of the decomposition of $r$ into factors. The
specific assumption we make is that the dimension of the hyperspin
space ${\mathbb S}^{A}$ is \emph{even}. Then we write $r=2n$, where
$n=1,2,\ldots$, and set ${\mathbb S}^{A}={\mathbb S}^{{\bf A}i}$,
where ${\bf A}$ is a two-component spinor index, and $i$ will be
called an `internal' index $(i=1,2, \ldots,n)$. Thus we can write
${\mathbb S}^{{\bf A}i}={\mathbb S}^{{\bf A}} \otimes{\mathbb
H}^{i}$, where ${\mathbb S}^{{\bf A}}$ is a standard spin space of
dimension two, and ${\mathbb H}^{i}$ is a complex vector space of
dimension $n$. The breaking of symmetry amounts to the fact that we
can identify the hyperspin space with the tensor product of these
two spaces.

We shall assume that ${\mathbb H}^{i}$ is endowed with a strong
Hermitian structure, i.e. that there is a canonical anti-linear
isomorphism between the complex conjugate of the internal space
${\mathbb H}^{i}$ and the dual space ${\mathbb H}_{i}$. If
$\psi^i\in{\mathbb H}^{i}$, then we write ${\bar\psi}_i$ for the
complex conjugate of $\psi^i$, where ${\bar\psi}_i\in{\mathbb
H}_{i}$. We see that ${\mathbb H}^{i}$ is a complex Hilbert
space---and indeed, although here we consider mainly the case for
which $n$ is finite, one should have in mind also the infinite
dimensional situation. For the other hyperspin spaces we write
${\mathbb S}_{A}={\mathbb S}_{{\bf A}i}$, ${\mathbb S}^{A'}={\mathbb
S}^{{\bf A}'}_{\ \ i}$, and ${\mathbb S}_{A'}={\mathbb S}_{{\bf
A}'}^{\ \ i}$, respectively. These identifications  preserve the
duality between ${\mathbb S}^{A}$ and ${\mathbb S}_{A}$, and between
${\mathbb S}^{A'}$ and ${\mathbb S}_{A'}$; and at the same time are
consistent with the complex conjugation relations between ${\mathbb
S}^{A}$ and ${\mathbb S}^{A'}$, and between ${\mathbb S}_{A}$ and
${\mathbb S}_{A'}$. Hence if $\alpha^{{\bf A}i} \in{\mathbb S}^{{\bf
A}i}$ then under complex conjugation we have $\alpha^{{\bf A}i}\to
{\bar\alpha}^{{\bf A}'}_{\ \ i}$, and if $\beta_{{\bf
A}i}\in{\mathbb S}_{{\bf A}i}$ then $\beta_{{\bf
A}i}\to{\bar\beta}_{{\bf A}'}^{\ \ i}$.

In the case of a quantum space-time vector $r^{\,AA'}$ we have a
corresponding structure induced by the identification
$r^{\,AA'}=r^{{\bf AA}'i}_{\ \ \ \ \ j}$. When the quantum
space-time vector is real, the weak Hermitian structure on
$r^{\,AA'}$ is manifested in the form of a weak Hermitian structure
on the two-component spinor index pair, together with a strong
Hermitian structure on the internal index pair. In other words, the
Hermitian condition on the space-time vector $r^{\,AA'}$ is given by
${\bar r}^{{\bf A}'{\bf A}i}_{\ \ \ \ \ j} = r^{{\bf AA}'i}_{\ \ \ \
\ j}$.

One consequence of these relations is that we can interpret each
point in a quantum space-time as being a
\emph{space-time-point-valued operator}. The ordinary classical
space-time then `sits' inside the quantum space-time in a canonical
manner---namely, as the locus of those points of quantum space-time
that factorise into the product of a space-time point $x^{{\bf
AA}'}$ and the identity operator on the internal space: $x^{{\bf
AA}'i}_{\ \ \ \ \ j} = x^{{\bf AA}'} \delta^{i}_{\,j}$. Thus, in
situations where special relativity is a satisfactory theory, we
regard the relevant physical events as taking place on or in the
immediate neighbourhood of the embedding of Minkowski space in
${\mathcal H}^{4n^2}$.

This picture can be presented in more geometric terms as follows. We
introduce the notion of a \emph{hypertwistor} as a pair of
hyperspinors $(\omega^A, \pi_{A'})$ with the pseudo-norm
$\omega^A{\bar\pi}_{A} + \pi_{A'}{\bar\omega}^A$. The projective
hypertwistor space ${\mathbb P}^{2r-1}$ in the case $r=2n$ admits a
Segr\'e embedding of the form ${\mathbb P}^3 \times {\mathbb
P}^{n-1}\subset{\mathbb P}^{4n-1}$. Many such embeddings are
possible, though they are all equivalent under the action of the
symmetry group $U(2n,2n)$. If the symmetry is broken and one such
embedding is selected out, then we can introduce homogeneous
coordinates and write $Z^{\alpha i}$ for the generic hypertwistor.
Here the Greek letter ${\alpha}$ denotes an `ordinary' twistor index
$({\alpha}=0,1,2,3)$ and $i$ denotes an internal index
$(i=1,2,\ldots,n)$. These two indices, when clumped together,
constitute a hypertwister index. The Segr\'e embedding consists of
those points in ${\mathbb P}^{4n-1}$ for which we have a product
decomposition given by $Z^{\alpha i}= Z^{\alpha}\psi^i$. Once
symmetry breaking takes place---and this may happen in stages,
corresponding to a successive factorisation of the underlying
hypertwistor space---then one can think of ordinary four-dimensional
space-time as becoming more or less disentangled from the rest of
the universe, and behaving to some extent autonomously. Nonetheless,
we might expect its global dynamics, on a cosmological scale, to be
affected by the distribution of mass and energy elsewhere in the
quantum space-time; and thus we obtain a possible model for `dark
energy'.

The idea of symmetry breaking being put forward here is related to
the notion of \emph{disentanglement} in quantum mechanics
\cite{brody1,gibbons}. That is to say, at the unified level the
degrees of freedom associated with space-time symmetry are quantum
mechanically entangled with the internal degrees of freedom
associated with microscopic physics. The phenomena responsible for
the breakdown of symmetry are thus analogous to the mechanisms of
decoherence through which quantum entanglements are gradually
diminished. There is also in this connection an interesting relation
to the so-called \emph{twistor internal symmetry groups} (see, e.g.,
\cite{hughston}).

Let us now examine the implications of our symmetry breaking
mechanism for fields defined on a quantum space-time. For example,
let $\phi(x^{\,AA'})$ be a scalar field on a quantum space-time.
After we break the symmetry by writing $x^{\,AA'}= x^{{\bf AA}'i}_{\
\ \ \ \ j}$, we consider an expansion of the field around the
embedded Minkowski space. More specifically, for such an expansion
we have
\begin{eqnarray}
\phi(x^{\,AA'}) = \phi^{(0)}(x^{{\bf AA}'}) + \phi^{(1)\ \ i}_{{\bf
AA}'\,j}(x^{\,{\bf AA}'})\left( x^{{\bf AA}'j}_{\ \ \ \ \ i} -
x^{{\bf AA}'}\delta^j_{\ i} \right) + \cdots,
\end{eqnarray}
where $\phi^{(0)}(x^{{\bf AA}'}) =\phi(x)|_{x=x^{{\bf AA}'}
\delta^{j}_{\,i}}$, and $\phi^{(1)\ \ i}_{{\bf AA}'\,j} (x^{\,{\bf
AA}'}) = (\partial\phi(x) / \partial x) |_{x=x^{{\bf AA}'}
\delta^{j}_{\,i}}$. Therefore the order-zero term defines a
classical field on Minkowski space, and the first-order term can be
interpreted as a `multiplet' of fields, transforming according to
the adjoint representation of $U(n)$. Alternatively, if the internal
space is infinite-dimensional, we can think of the first-order term
as a field operator.

\begin{figure}
  \includegraphics[height=.4\textheight,angle=270]{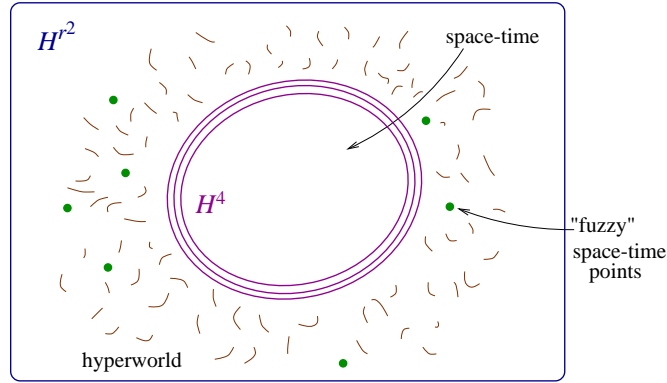}
  \caption{\label{fig:2}{\sl Fuzzy space-time}. Once the
  symmetry is broken, a quantum space-time event can be
  interpreted as a space-time-point-valued operator, whose
  expectation determines a point in Minkowski space.}
\end{figure}

\section{Emergence of quantum states}

The embedding of Minkowski space in the quantum space-time
${\mathcal H}^{4n^2}$ implies a corresponding embedding of the
Poincar\'e group in the hyper-Poincar\'e group. This can be seen as
follows. The standard Poincar\'e group in ${\mathcal H}^{4}$
consists of transformations of the form $x^{{\bf AA}'}
\longrightarrow\, \lambda^{\bf A}_{\bf B}\, \bar{\lambda}^{{\bf
A}'}_{{\bf B}'}\, x^{{\bf BB}'} + \beta^{{\bf AA}'}$. This action
lifts naturally to a corresponding action on ${\mathcal H}^{4n^2}$
given by $x^{{\bf AA}'i}_{\ \ \ \ \ j}\longrightarrow\, \lambda^{\bf
A}_{\bf B}\, \bar{\lambda}^{{\bf A}'}_{{\bf B}'}\, x^{{\bf BB}'i}_{\
\ \ \ \ j} + \beta^{{\bf AA}'}\delta^i_{\,j}$. On the other hand,
the general hyper-Poincar\'e transformation in the broken symmetry
phase can be expressed in the form
\begin{eqnarray}
x^{{\bf AA}'i}_{\ \ \ \ \ j}\longrightarrow\, L^{{\bf A}i}_{{\bf
B}k}\, \bar{L}^{{\bf A}'l}_{{\bf B}'j}\, x^{{\bf BB}'k}_{\ \ \ \ \
l} + \beta^{{\bf AA}'i}_{\ \ \ \ \ j}. \label{eq:12.3}
\end{eqnarray}
Thus the embedding of the Poincar\'e group as a subgroup of the
hyper-Poincar\'e group is given by $L^{{\bf A}i}_{{\bf B}j} =
\lambda^{\bf A}_{\bf B}\delta^i_{\,j}$ and $\beta^{{\bf AA}'i}_{\ \
\ \ \ j} = \beta^{{\bf AA}'} \delta^i_{\,j}$.

Bearing this in mind, we construct a class of maps from the quantum
space-time ${\mathcal H}^{4n^2}$ to Minkowski space ${\mathcal
H}^{4}$. Under rather general physical assumptions, such maps are
necessarily of the form
\begin{eqnarray}
x^{{\bf AA}'i}_{\ \ \ \ \ j}\longrightarrow\, x^{{\bf AA}'} =
\rho^j_i\,x^{{\bf AA}'i}_{\ \ \ \ \ j}, \label{eq:12.5}
\end{eqnarray}
where $\rho^j_i$ is a \emph{density matrix}. By a density matrix we
mean, as usual, a positive semi-definite strongly Hermitian matrix
with unit trace. The maps thus arising here can be regarded as
quantum expectations. In particular, let $\rho:\ {\mathcal
H}^{4n^2}\to {\mathcal H}^4$ satisfy the following conditions: {\rm
(i)} $\rho$ is linear and maps the origin of ${\mathcal H}^{4n^2}$
to the origin of ${\mathcal H}^{4}$; {\rm (ii)} $\rho$ is Poincar\'e
invariant; and {\rm (iii)} $\rho$ preserves causal relations. Then
$\rho$ is given by a density matrix on the internal space (we refer
the reader to \cite{brody2} for a proof).

This result shows how the causal structure of quantum space-time is
linked with the probabilistic structure of quantum mechanics. The
concept of a quantum state emerges when we ask for consistent ways
of `averaging' over the geometry of quantum space-time in order to
obtain a reduced description of physical phenomena in terms of the
geometry of Minkowski space. We see that a probabilistic
interpretation of the map from a general quantum space-time to
Minkowski space arises as a consequence of elementary causality
requirements. We can thus view the space-time events in ${\mathcal
H}^{4n^2}$ as representing space-time-point-valued quantum
observables, the totality of which constitute a `fuzzy space-time'
(see Figure~\ref{fig:2}), and the expectations of these fuzzy
space-time points correspond to points of ${\mathcal H}^{4}$.

\begin{figure}
  \includegraphics[height=.19\textheight,angle=00]{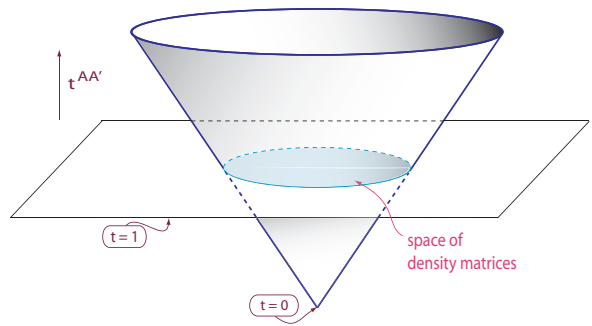}
  \caption{\label{fig:3}{\sl Space of density matrices}. The
  intersection of the cone of positive Hermitian matrices with a
  space-like hypersurface is the space of density matrices. }
\end{figure}

Finally, we remark that the space of density matrices itself is
endowed with a natural Finslerian metric induced from the ambient
pseudo-Finslerian structure, as illustrated in Figure~\ref{fig:3}
(cf. \cite{wojtkowski}).

\end{document}